\documentclass[traditabstract]{aa}
\usepackage[varg]{txfonts}
\usepackage{graphicx}
\usepackage{xcolor}
\usepackage{natbib}
\usepackage{amsmath,amssymb}
\usepackage{hyperref}
\usepackage{rotating, bm}

\newcommand{\ket}[1]{\left|#1\right\rangle}
\newcommand{\sref}[1]{Sect.\,\ref{#1}}
\newcommand{\fref}[1]{Fig.\,\ref{#1}}
\newcommand{\eref}[1]{Eq.\,(\ref{#1})}

\begin{document}
\titlerunning{Radiation transfer effects in He\,{\sc i}\,10830}
\title{The He\,{\sc i}\,10830\,\AA\ line: Radiative transfer and differential illumination effects}

\author{Andr\'es Vicente Ar\'evalo\inst{1,2}
\and Ji\v{r}\'{\i} \v{S}t\v{e}p\'an\inst{3}
\and Tanaus\'u del Pino Alem\'an\inst{1,2}
\and Mar\'ia Jes\'us Mart\'inez Gonz\'alez\inst{1,2}}

\authorrunning{Vicente Ar\'evalo et al.}

\institute{Instituto de Astrof\'{\i}sica de Canarias, E-38205 La Laguna, Tenerife, Spain
\and Departamento de Astrof\'{\i}sica, Universidad de La Laguna, E-38206 La Laguna, Tenerife, Spain
\and Astronomical Institute of the Academy of Sciences, Ond\v{r}ejov, Czech Republic.
}

\date{Received XXXX; accepted XXXX}

\abstract{
We study the formation of the Stokes profiles of the \ion{He}{i} multiplet at 10830\,\AA\ when
relaxing two of the approximations that are often considered in the modeling of this multiplet,
namely the lack of self-consistent radiation transfer and the assumption of equal
illumination of the individual multiplet components. This \ion{He}{i} multiplet is among
the most important ones for the diagnostic of the outer solar atmosphere from
spectropolarimetric observations, especially in prominences, filaments, and spicules.
However, the goodness of these approximations is yet to be assessed, especially in situations 
where the optical thickness is of the order or larger than one, and radiation
transfer has a significant impact in the local anisotropy and the ensuing spectral line
polarization.
This issue becomes particularly relevant
in the ongoing development of new inversion tools which take into account multi-dimensional
radiation transfer effects. To relax these approximations we generalize the multi-term
equations for the atomic statistical equilibrium to allow for differential illumination
of the multiplet components and implement them in a one-dimensional radiative transfer
code. We find that, even for this simple geometry and relatively small optical thickness, both radiation transfer and differential illumination effects have a significant
impact on the emergent polarization profiles. This should be taken into account in order to
avoid potentially significant errors in the inference of the magnetic field vector.
}

\keywords{
Atomic processes --
Polarization --
Radiative transfer --
Sun: atmosphere
}

\maketitle
 
\section{Introduction\label{sec:intro}}

The magnetic field is fundamental to understand commonly observed plasma structures
such as prominences, filaments, and spicules, being responsible for their
structure, properties, and even their existence (e.g., the reviews of
\citealt{2010SSRv..151..333M} and \citealt{Tsiropoulaetal2012}).
Spectropolarimetric observations of the \ion{He}{i} multiplets at 10830\,\AA\
(hereafter, 10830 multiplet) and 5876\,\AA\ (usually dubbed D$_3$) have been
extensively acquired and analyzed to diagnose these structures and, in particular,
to infer their magnetic fields \citep[see][and references therein]{2022ARA&A..60..415T}.

The formation of the Stokes profiles of these ortohelium spectral lines is
somewhat difficult to model due to their sensitivity to the coronal UV/FUV
illumination \citep{2008ApJ...687.1388J},
which ionizes the neutral helium atoms that then can be recombined to
populate the relatively high excitation energy triplet states. However,
this property is what has also allowed to significantly simplify the
modeling of these lines. It turns out that the 10830 multiplet and D$_3$
cannot effectively form in quiet Sun conditions, being only observable in magnetically
active regions and in plasma structures such as prominences, filaments,
and spicules. Because they form in more or less localized plasma regions
with somewhat small optical thickness, it is generally
assumed that it is possible to model these multiplets with a relatively simple slab model, without
accounting for radiation transfer (RT) effects and with the most complex processes
such as the coronal illumination abstracted into the optical depth of the slab.
This fact has been exploited in the HAZEL inversion code
\citep{2008ApJ...683..542A} that has been widely used for the analysis
of spectropolarimetric observations in prominences, filaments, and
spicules \citep[see the review][and references therein]{2022ARA&A..60..415T}.

Apart from this apparent simplicity in their modeling, the 10830 and D$_3$
multiplets are both well observable with today's instrumentation and their
spectral line polarization is sensitive to the magnetic field with
strengths between a fraction of a gauss to some hundred gauss,
values expected to be typical in the outer atmosphere structures.
Moreover, their polarization is sensitive to the magnetic field via the Hanle and Zeeman effects,
and elastic collisions with neutral hydrogen atoms in chromospheric and prominence plasma are unable
to destroy the atomic polarization of the \ion{He}{i} levels \citep{2009ApJ...701L..43C}.
All these facts have made these \ion{He}{i} multiplets really useful for the inference of the magnetic field vector in the above mentioned regions of the solar atmosphere.

The formation of the Stokes profiles of the 10830 and D$_3$ multiplets can
be described with the quantum theory of atomic line formation \citep{LL04}.
In particular, the multi-term model atom is the most suitable for this
application \citep[see sections 7.5, 7.6, and 13.4 of][for a detailed description of the problem]{LL04}.
One important requirement for the applicability of the multi-term model
atom is that the exciting radiation field must be spectrally flat\footnote{
The Stokes parameters are constant with wavelength.} over the wavelength range
spanned by the multiplet components. For the 10830 multiplet this implies
that the blue and red components (remember that the red component is a blend
of two of the lines of this triplet), which are about 1\,\AA\ apart, must be
excited by identical radiation fields (and likewise for D$_3$ and its components).

This assumption is very well satisfied if the optical thickness of the
multiplet components is small ($\tau<1$) and the exciting illumination
is that of the relatively flat continuum of the quiet photosphere.
However, observations clearly
indicate that the optical thickness of the 10830 multiplet can often
exceed one \cite[e.g.][]{2019A&A...625A.128D,2019A&A...625A.129D}.
For optically thick enough plasmas, RT effects within
the region of formation of the \ion{He}{i} multiplets lead to a
spectrally non-flat radiation field. Due to the non-negligible separation
in wavelength between the red and blue components of the 10830 multiplet,
and the difference of their optical thicknesses, the radiation field becomes
indeed non-flat and the multi-term model atom equations are no longer
suitable. Even though the potential importance of these RT
effects in the 10830 multiplet have been recognized before
\citep[see][]{2007ApJ...655..642T}, no detailed investigation of this
problem has ever been conducted.

The so-called flat-spectrum condition or approximation is due to the fact that the theory of complete frequency redistribution \citep[CRD,][]{LL04} is based on the first-order
perturbative expansion of the matter--radiation interaction. In order to relax this condition, it is necessary to use a higher-order theory, which allows considering coherent scattering processes and partial frequency redistribution (PRD) effects 
(\citealt{Stenflo1994, Bommier1997,Bommier1997b,Casinietal2014}, or \citealt{Bommier2017}).
However, including PRD effects dramatically increases the computing time
requirements, making it not the most desirable approach for a
multiplet that can be successfully modeled by assuming complete
frequency redistribution \citep{2008ApJ...683..542A}. Assuming
a multi-level model atom \citep[see sections 7.1 and 7.2 of][]{LL04} also
naturally relaxes this assumption. However, the quantum interference between the
upper levels of the blended red components of the 10830 multiplet needs to
be taken into account to correctly model their polarization, which is not possible
within this model.

In this paper, we propose a new approach to the 10830 multiplet formation that is more general than the multi-term model atom, at least for the magnetic fields strengths relevant for chromospheric and coronal spectropolarimetry.
Our formulation allows us to treat separately the illumination of the red and blue components of the 10830 multiplet.
The multi-term model atom is the limit case of our method, strictly valid in case of spectrally flat illumination and negligible optical thickness of the medium. In contrast to the multi-term approximation, our approach allows to solve the problems out of the local thermodynamic equilibrium approximation (NLTE) in plasmas of any optical thickness.
As we show below, this approach leads to significant modification of the traditional results in 1D slab models.
Moreover, NLTE RT plays an even more important role in the formation of
the lines of the outer solar atmosphere if 3D effects are considered
\citep{2022A&A...659A.137S}

In order to be able to consider NLTE models involving the 10830 line, we need to realize
that the quantum interference between the upper level of the blue component and the other two
levels in the term is not expected to have a significant impact for the
typical magnetic fields found in the solar atmosphere. We thus derive
a new set of statistical equilibrium equations (SEE) starting from the multi-term
model atom of \cite{LL04} and explicitly removing the quantum interference
between the upper level of the blue component and the rest of the levels, what
allows us to introduce different pumping radiation fields for the blue and
red components of the 10830 multiplet. We have implemented this new set
of equations into a one-dimensional RT code. In
\sref{sec:formul}, we describe the new set of SEE and some details
about the RT code. In \sref{sec:tests}, we
carry out a series of numerical experiments
to study the impact of RT and differential illumination effects
on the emergent Stokes profiles. Finally, we present our conclusions in
\sref{sec:concl}.

\section{Formulation of the problem\label{sec:formul}}

\begin{figure*}[htp!]
\begin{center}
\includegraphics[width=0.49\textwidth]{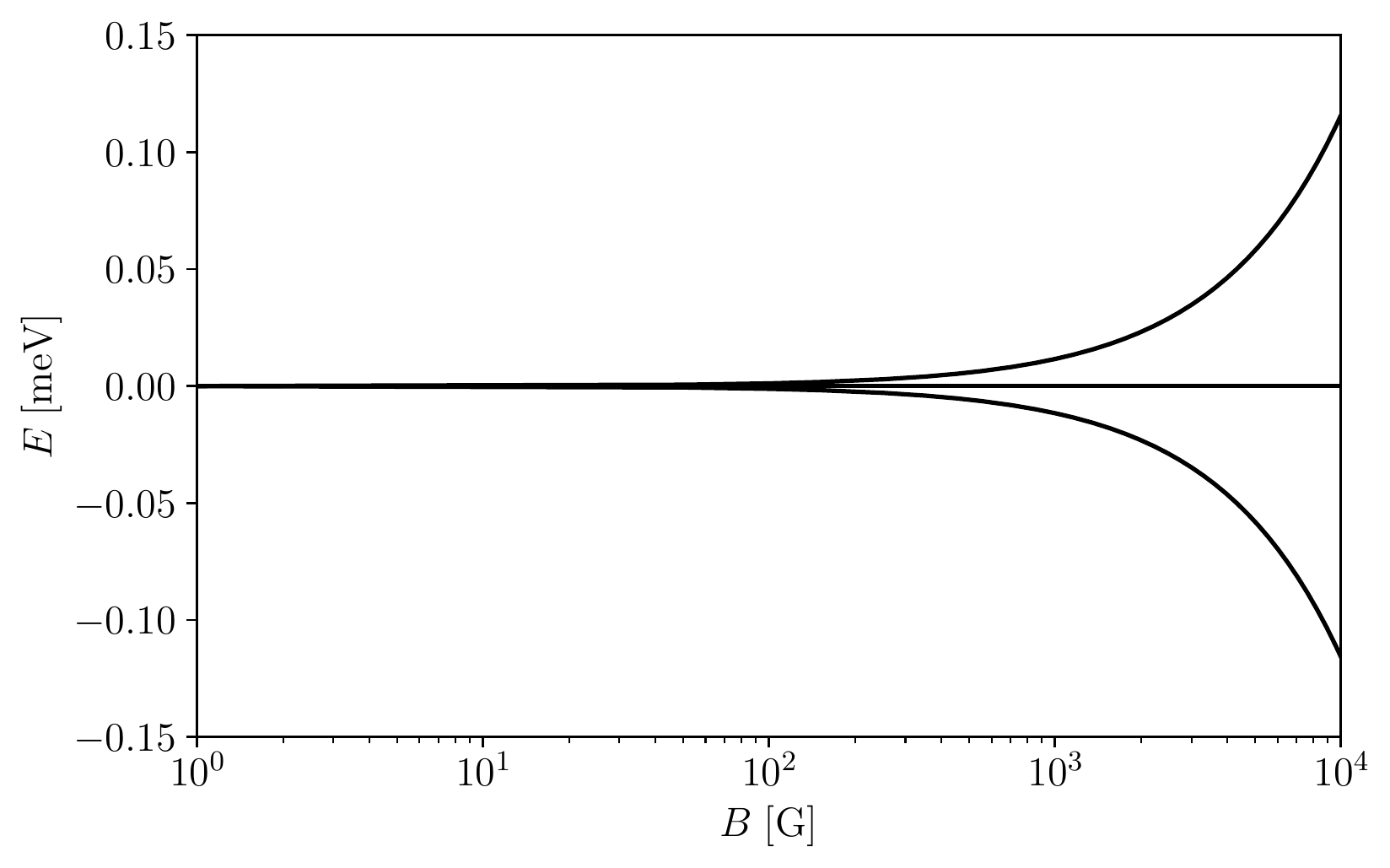}
\includegraphics[width=0.49\textwidth]{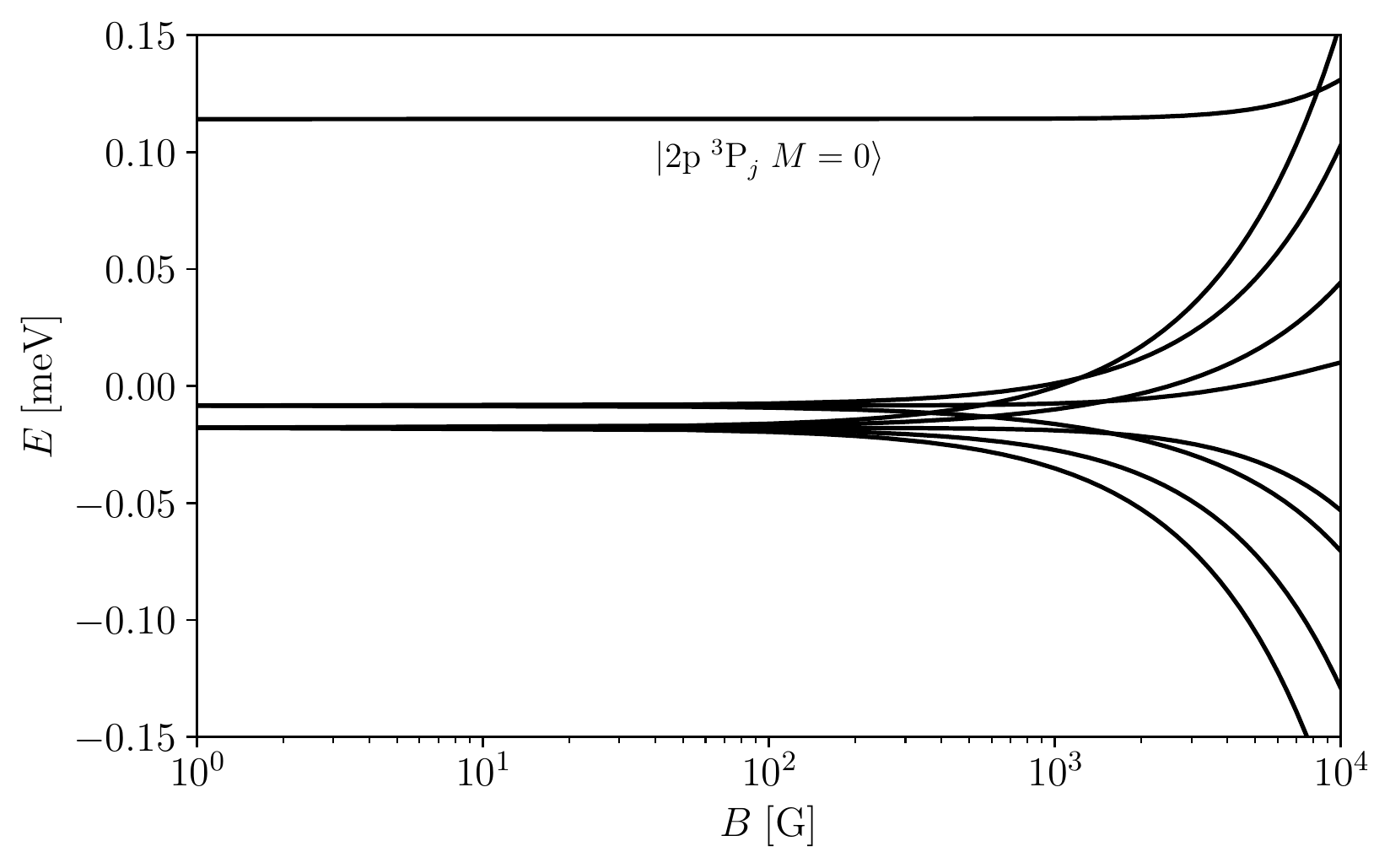}
\end{center}
\caption{Energy of the magnetic sublevels of the lower (left panel) and upper (right panel)
terms of the 10830\,\AA\ multiplet as a function of the magnetic field strength. The natural
widths of the upper-term levels is about $10^{-5}$\,meV, well below the plotting resolution. The zero energy offsets in each panel correspond to the mean energies of the respective terms.
}
\label{fig:paschen}
\end{figure*}

The theory of atomic line polarization summarized in the monograph by \citet{LL04} is
formulated in the frame of the so-called complete frequency redistribution (CRD). This
limit of atom-photon interactions, which implies a complete lack of correlation between
the frequencies of the absorbed and emitted photons in scattering processes, has been
immensely useful for inferring magnetic fields of the solar atmosphere during the last
decades.

The multi-term model atom, described in Chapter\,7 of \citet{LL04} is the most
suitable to describe the 10830 multiplet as it accounts for quantum interference between
states $\ket{\beta LSJM}$ and $\ket{\beta LSJ'M'}$ of different $J$ and $J'$ levels of
the same $\beta LS$ term. However, in order to ensure physical consistency when
accounting for quantum interference between non-degenerate atomic levels, the incident
radiation field must have a flat spectrum across a frequency range wider than the
separation of those levels.
In addition, the CRD theory is strictly valid if the incident radiation field is flat on a
frequency interval much larger than the natural width of the atomic states. Due to the small
natural width of the 10830 multiplet sublevels, this condition is automatically satisfied
if the spectrum is flat across the whole multiplet.

When these conditions are satisfied, the absorption and stimulated emission within a
spectral line depend on the frequency independent radiation field tensor
\begin{align}
\overline{J}^K_{Q}=\int J^K_{Q}(\nu)\phi(\nu)\;d\nu\,,
\label{eq:jkq}
\end{align}
where $\phi(\nu)$ is the normalized line's absorption profile, and $J^K_Q(\nu)$ is the radiation
field tensor at each frequency $\nu$. Note that the $\phi(\nu)$ absorption profile is representative of the absorption in the whole multiplet and thus it has the shape of the absorptivity, and not that of a Voigt profile. Strictly speaking, this approach
is only valid if all the multiplet components have comparable absorptivities and are illuminated by similar enough radiation fields, what is not necessarily true for the 10830 multiplet.

It is then clear that, if we want to introduce different radiation fields for the blue
and red components in the SEE, we need to neglect quantum
interference between the blue component's upper level and any of the red component's upper
levels. Consequently, only the two red component's upper levels can be coherent. In this
way we can still fulfill the validity condition of the CRD approximation, namely spectral
flatness on a frequency interval much larger than the natural width of each transition,
as well as the validity condition of the multi-term atom model, namely spectral flatness
in a frequency range wider than the separation between levels that can be coherent.

Generally, the further in energy two levels are, the less significant quantum interference
between them is. The separation between the $J=0$ state, the upper level of the blue
component of the 10830 multiplet, from the rest of the states is about $30$\,GHz or
$0.1\,\mathrm{meV}$, i.e., about four orders of magnitude larger than the natural width of
the Zeeman states, which is about $1.6\,\mathrm{MHz}$ or $10^{-5}\,\mathrm{meV}$. This energy
separation remains very large even if the magnetic states are modified by a magnetic field with
strength up to $\sim5$\,kG, when some of the Zeeman components of the upper levels of the
red component crosses with the blue component's upper level (see \fref{fig:paschen}).
Therefore, for the typical magnetic fields found in the solar atmosphere, we can safely
neglect quantum interference between the upper levels of the blue and red components, and thus
we can consider different radiation fields for each component while ensuring that the
CRD theory remains strictly valid.

Consequently, the only difference with respect to the standard multi-term SEE (see section
7.6 in \citealt{LL04}) is that, instead of considering a single radiation field for the
whole multiplet, we distinguish between the radiation field in the red and blue components,
forcing the quantum interference between magnetic states pertaining to different components
to vanish.
Therefore, instead of a single $\overline{J}^K_Q$ radiation tensor common to all the multiplet sublevels, we now have $\overline{J}^K_Q({\rm red})$ or $\overline{J}^K_Q({\rm blue})$, resulting from the same average as in Eq.\,\eqref{eq:jkq}, but integrating over $\phi(\nu)_{\rm red}$ and $\phi(\nu)_{\rm blue}$--the absorption profiles accounting only for the contributions for the red or the blue component--, respectively.
In our approach we follow the standard way of derivation of the equations and we diagonalize the atomic Hamiltonian in the
incomplete Paschen-Back effect regime.
Consequently, the RT coefficients in the RT equation 
have exactly the same formal expression as the corresponding coefficients of \citet{LL04}.
These relatively minor changes in the SEE allow us to consider a much broader set of physical
scenarios.

We have implemented these SEE in a 1D RT code, which solves the NLTE problem
of the generation and transfer of polarized radiation.

\section{Numerical experiments
\label{sec:tests}}

In this section we show the result of some numerical experiments
to illustrate how accounting for differential radiation between the red and
blue components of the 10830 multiplet, as well as RT effects,
can lead to strikingly different emergent Stokes profiles.

We compare the results obtained with our code with those obtained under the
assumption of flat-spectrum and 
negligible impact of the RT effects on the atomic density matrix.
For this physical scenario, the RT
equations have the solution \citep[see, e.g.,][]{2008ApJ...683..542A}:
\begin{align}
    {\bm I} = \left[{\bm 1} + \psi_{\rm O} {\bm K}'\right]^{-1} \left[ (\mathrm{e}^{-\tau}{\bm 1} - \psi_{\rm M}{\bm K}'){\bm I}_\mathrm{inc} + (\psi_{\rm M} + \psi_{\rm O}){\bm S} \right]\,, 
    \label{eq:analytical_rt}
\end{align}
where ${\bm 1}$ is the unit matrix, ${\bm K}'={\bm K}/\eta_I-{\bm 1}$, with ${\bm K}$ the propagation
matrix and $\eta_I$ the absorption coefficient for intensity, ${\bm S}$ is the source function vector,
and ${\bm I}_\mathrm{inc}$ is the Stokes vector of the incident radiation (at the lower boundary of
the slab). The coefficients $\psi_{\rm O}$ and $\psi_{\rm M}$ only depend on the optical thickness
along the propagation direction at a particular frequency and angle, and their expression can be found
in \cite{KunaszAuer1988}. Due to its simplicity and straightforward evaluation, the linear
\eref{eq:analytical_rt} is commonly used in practical applications to invert spectropolarimetric data
of the 10830 multiplet.

As noted above, Eq.\,\eqref{eq:analytical_rt} is applicable if both ${\bm K'}$ and ${\bm S}$ are constant along the ray of propagation. This is a good approximation if the optical
thickness is below unity. Note that, while this approximation does include radiation
transfer via Eq.\,\eqref{eq:analytical_rt}, it is not self-consistent, as the radiation
field is assumed fixed and constant throughout the whole extension of the slab.
However, for larger optical thicknesses this approximation becomes unsuitable because RT
starts playing a significant role \citep[see][]{2007ApJ...655..642T}. The problem
then becomes non-local and non-linear, and the notably different opacities between the red
and blue components also lead to the non-fulfillment of the spectral flatness approximation.

Equation\,\eqref{eq:analytical_rt} is strictly valid for an optically thin slab illuminated with a spectrally flat incident radiation. We have checked that, in this limit, our calculations coincide with the result of Eq.\,(\ref{eq:analytical_rt}) for magnetic fields from zero to several thousands of gauss (see \sref{ssec:profiles}).

\subsection{Impact on the radiation field anisotropy\label{ssec:slab2}}

\begin{figure*}
\begin{center}
\includegraphics[width=\textwidth]{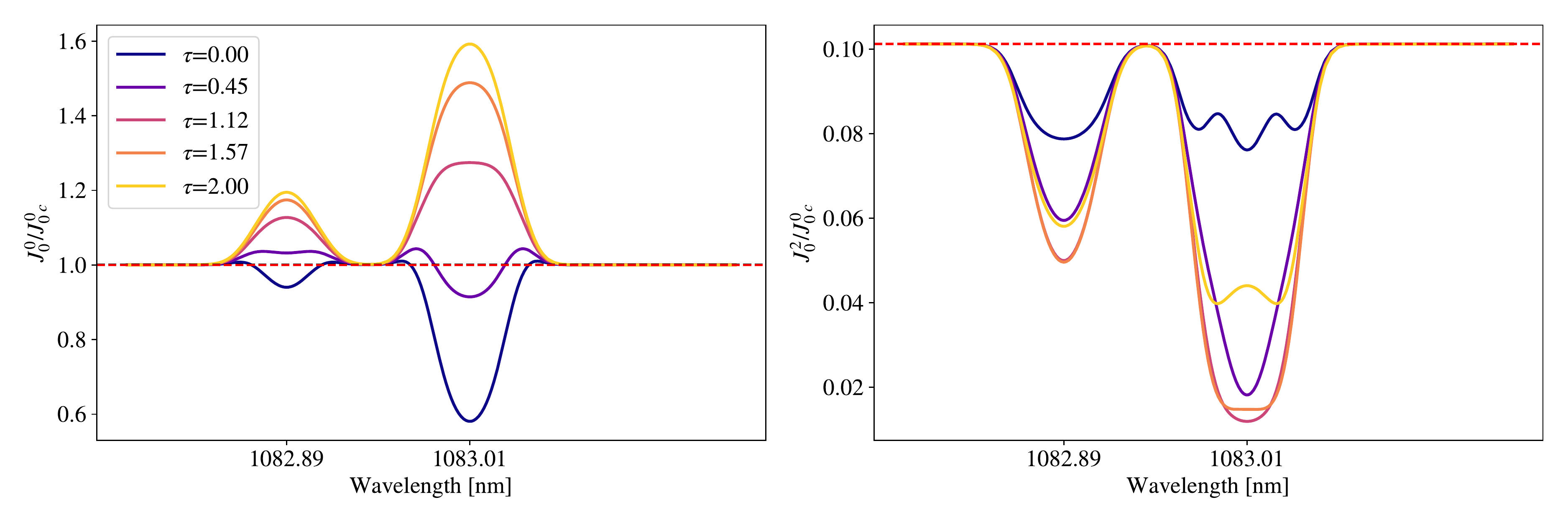}
\end{center}
\caption{
Mean intensity ($J^0_0$; left panel) and anisotropy ($J^2_0$; right panel) normalized to the
incident continuum mean intensity ($J^0_0{}_\mathrm{c}$) at different optical depths $\tau$,
measured from the top boundary at the 10830 multiplet red component's center. Color coded we have different
optical depth layers of the fully consistent NLTE solution. The red dashed line corresponds to the
incident radiation field (${\bm I}_\mathrm{inc}$ in Eq.\,\ref{eq:analytical_rt}).}
\label{fig:radiation}
\end{figure*}

In this experiment, we consider a slab with constant properties located 6\,Mm above the
solar surface, with its normal axis parallel to the solar radius. The optical thickness
of the slab is $\tau=2$ in the center of the red component of the 10830 multiplet. We
solve the self-consistent RT transfer problem in this slab model.

In this first experiment we assume that the slab is unmagnetized (${\bm B}=0$), and we
analyze the wavelength variation of the $J^0_0$ and $J^2_0$ (mean intensity and anisotropy,
respectively) components of the radiation field tensor at different optical depths from
the top, $\tau=0$, to the bottom, $\tau=2$, boundaries (see \fref{fig:radiation}). In
Eq.\,\eqref{eq:analytical_rt}, the radiation field tensor is constant throughout the
slab and spectrally flat (red dashed line, hereafter non-RT model).
However, the radiation field tensor components in the self-consistent solution (solid curves,
hereafter RT model) show a strong dependence with height due to the combination of RT
effects and the differential absorption and emission in the red and blue components of the
multiplet.

At the top of the slab ($\tau=0$, dark blue curve) the mean intensity spectrum (left panel of
\fref{fig:radiation}) resembles the typical 10830 absorption profiles. This is easily understood
mostly in terms of the absorption of the incident intensity in the slab's bottom boundary. At the
bottom of the slab ($\tau=2$, yellow curve), however, the mean intensity shows an excess with
respect to the mean intensity of the incident continuum radiation. This is due to the radiation
that is emitted from within the slab, traveling ``downward''.

Regarding the radiation field anisotropy (right panel of \fref{fig:radiation}), we see a reduction
with respect to the incident anisotropy, a reduction which is non-monotonic with the optical depth.
This reduction is mainly a consequence of the horizontal RT within the slab: for inclined
lines of sight, there is a larger amount of emitting material, which increases the negative
contribution of the radiation coming from directions forming an angle with the slab's axis larger
than the van~Vleck angle \citep[see, e.g.,][]{Trujillo2001}. For this reason, the largest differences
between the non-RT and RT model's anisotropies are found at optical depths around $\tau=1$, where
the radiation is more likely to start escaping the slab along the vertical direction, while the more
inclined rays are still optically thick.

From this experiment, it is clear that RT and the relaxation of the flat spectrum approximation
can have a considerable impact on the radiation field tensors that, as we will show in subsequent
experiments, can significantly impact the emergent Stokes parameters.

\subsection{Impact on the emergent Stokes profiles\label{ssec:profiles}}

Even with a relatively small optical thickness of $\tau=2$, NLTE RT effects significantly
modify the radiation field anisotropy within the slab. Since this quantity is crucial for
the emitted linear scattering polarization we can expect RT effects to have a significant
impact on the emergent Stokes profiles as well.

\begin{figure*}
\begin{center}
\includegraphics[width=\textwidth]{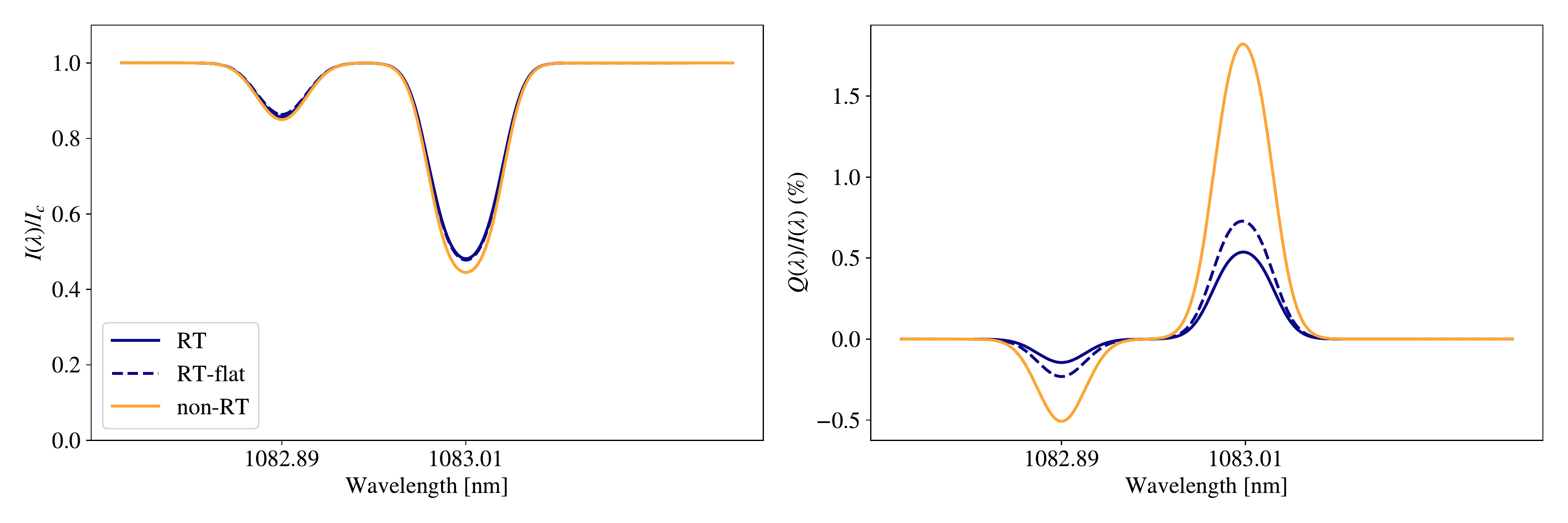}
\end{center}
\caption{Intensity $I$ (left panel) and $Q$ (right panel) line profiles normalized to the continuum intensity for both the constant-property slab model solution (orange) and fully consistent NLTE solution (blue) for disk-center observation of the $\tau = 2$ slab with horizontal magnetic filed with $B=10$\,G. The positive $Q$ direction is parallel to the projection of the magnetic field vector onto the plane of the sky.}
\label{fig:NLTE_semianalytical}
\end{figure*}

We consider the same physical scenario as in Sect.\,\ref{ssec:slab2}, but with a uniform
and horizontal (perpendicular to the slab's axis) magnetic field of $B=10$\,G. After solving
the self-consistent NLTE problem, we calculate the emergent Stokes profiles for a line of
sight parallel to the slab's axis. Note that if the slab and the incident illumination at
the bottom boundary are axially symmetric, as it would be the case if it was not for the
chosen magnetic field, there would be no polarization when observing along the slab's axis
due to the symmetry; the magnetic field is thus necessary to generate scattering linear
polarization via the Hanle effect in the chosen model and line of sight. This would
correspond, for example, to the observation of a filament close to the disk-center.
By choosing the reference direction for the linear polarization parallel to the magnetic
field vector, the only non-zero Stokes parameters are $I$ and $Q$.

In \fref{fig:NLTE_semianalytical} we show the emergent Stokes profiles for three cases: 
i) the non-RT model (orange solid curve), ii) the RT model with the flat spectrum across
the whole multiplet approximation (dashed blue curve, hereafter RT-flat model), and iii)
the RT model (solid blue curve).

Regarding the intensity (left panel of \fref{fig:NLTE_semianalytical}), the impact of the
approximations is relatively minor. Assuming or not flat spectrum across the whole multiplet
does not have a significant impact if one solves the self-consistent problem, and fully neglecting
RT effects only has some impact on the depth of the red component. This difference could
lead to a slight error on the determination of the optical depth or the Doppler width (equivalently,
the temperature) of the slab, but we do not expect such a difference to be significant.

More interesting is what happens to the polarization (right panel of \fref{fig:NLTE_semianalytical}).
The flat spectrum approximation in the self-consistent (RT-flat model) solution results
in a significant increase of the emergent linear polarization, changing the ratio between
the polarization signals of the blue and red components. Moreover, if RT is also neglected
(non-RT model), the calculated emergent linear polarization is even larger, resulting in a
more than a factor theree increase with respect to the RT model.

Consequently, both RT effects and the flat-spectrum approximation have a significant
impact on the linear polarization profiles. This difference in the signal is directly
related to the changes in the frequency independent anisotropy as defined in Eq.\,\eqref{eq:jkq}
with the suitable absorption profiles. In particular, the anisotropy in the red
component, which determines the alignment of the red component's upper levels, is smaller
in the RT model than in the non-RT model in the whole slab, and smaller than the anisotropy
of the RT-flat model in the upper part of the slab ($\tau\le1$) from where most of the
photons emerge. Curiously, the anisotropy of the blue
component is significantly larger than that of the RT-flat model and of the red component
in the RT model (although still smaller than the anisotropy in the non-RT model).
However, the blue component's upper level is non-polarizable ($J=0$) and thus
the polarization of this line is fully due to dichroism, that is, due to the atomic alignment
of the lower level \citep{TrujilloBuenoetal2002}. It turns out that the impact of the red component on the lower
level alignment is such that it is smaller when relaxing the flat spectrum approximation,
and therefore the emergent linear polarization in the RT model is consistently smaller across the
whole multiplet.

We can further study the impact of these approximations by comparing the fractional linear
polarization signal at the peaks of the blue and red components for different magnetic field
strengths and for several optical depths (see \fref{fig:NLTE_semianalytical}). For the chosen
model (disk-center line of sight, axially symmetric slab model) the horizontal magnetic field
breaks the axial symmetry and induces linear polarization. Up to a few gauss, in the Hanle regime, the
polarization increases with the magnetic field. For larger magnetic field strengths, the Hanle effect
is in saturation and the linear polarization is insensitive to changes in the magnetic field strength
(plateau in the linear polarization in \fref{fig:NLTE_semianalytical}). For magnetic field
strenghts in the hundreds the Zeeman effect starts affecting the linear polarization (end of
the plateau toward the largest field strengths in \fref{fig:NLTE_semianalytical}). For small
optical depths (e.g., $\tau=0.01$, optically thin limit, see top-left panel) the three
approaches produce, as expected, the same polarization signal for every magnetic field.
However, for increasingly larger optical depths, RT effects induce a significant reduction
of the anisotropy and thus of the polarization signal, which is more significant the larger
is the optical depth. This behavior will saturate when the optical depth is large enough as to
thermalize the bottom boundary of the slab. The relaxation of the flat-spectrum approximation
also has an impact on the emergent linear polarization signals (compare the solid and dashed
curves in \fref{fig:NLTE_semianalytical}), but the most significant reduction in the polarization
emergent from the slab is due to RT effects (compare the dashed and dotted curves in the figure).

In summary, the NLTE solution leads, in this particular model, to a significant decrease of
anisotropy within the slab and to the decrease of the 10830 polarization amplitude.
Consequently, the interpretation of the observations based on the optically thin slab model with
the flat-spectrum approximation, the non-RT model,
could lead to significant errors in the determination of the slab's physical properties, especially
the magnetic field vector. This error can be critical depending on the particular
physical scenario: assume an observation of a solar filament with a horizontal magnetic field in the
Hanle saturation regime (e.g., 20 gauss), for which we are able to delimit its height (if not,
we would face a different problem in the almost complete degeneration between height and magnetic
field strength as inversion parameters). The non-RT model would overestimate the anisotropy and the 
inversion algorithm
would need to pick a magnetic field strength in the Hanle regime, where the linear polarization is
still sensitive to the magnetic field strength, in order to find a smaller polarization signal that
fits the observation for the overestimated anisotropy. We would then infer magnetic
field strengths in the fractions of gauss ($\mathcal{O}(10^{-1})$), instead of a magnetic field
in the saturation regime ($\mathcal{O}(10^1$--$10^2)$).
The opposite would happen in the prominence scenario, where a horizontal field depolarizes the
zero-field scattering signal. In order to compensate for the excess in anisotropy from the non-RT
assumption, the inferred magnetic field is increased, what could lead to the identification of field
strengths of fractions of gauss ($\mathcal{O}(10^{-1})$) as magnetic fields in the saturation regime
($\mathcal{O}(10^1$--$10^2)$). We emphasize, however, that a non-RT model would also be unable to 
correctly fit the linear polarization profiles altogether. In the non-RT model the ratio between the
amplitudes of the red and blue components is fixed for each value of the signal of any of the two
components. Because the ratio between these two signals with RT is different, there is not
a combination of parameters (in a constant property slab) such that the emergent Stokes parameters
from Eq.\,\eqref{eq:analytical_rt} fit the RT profiles.

Although our conclusions are not model dependent,\footnote{The very same argument can
be made just by knowing that RT effects reduce the anisotropy, a fact that has been known for decades
\citep{2007ApJ...655..642T}.}
their quantification is. While the non-RT slab model shows these issues due to the non-RT assumption,
the RT slab model finds itself in the opposite extreme, that is, where the optical depths tends to
infinity in the horizontal direction. We must thus emphasize that, while our modeling exposes a
potential problem in the inference of magnetic fields with the non-RT model, what we show in this
paper is the worst case scenario. First, the reduction of the anisotropy calculated in
the slab is an upper limit. Secondly, the observation of circular polarization can alleviate the
problem by providing a constrain in the magnetic field longitudinal component (we have intentionally
chosen a physical scenario without circular polarization in this paper), even though this cannot
fully solve issues related to the direction of the magnetic field vector and it can even make it
impossible to fit the four Stokes parameters simultaneously.

\begin{figure*}
\begin{center}
\includegraphics[width=\textwidth]{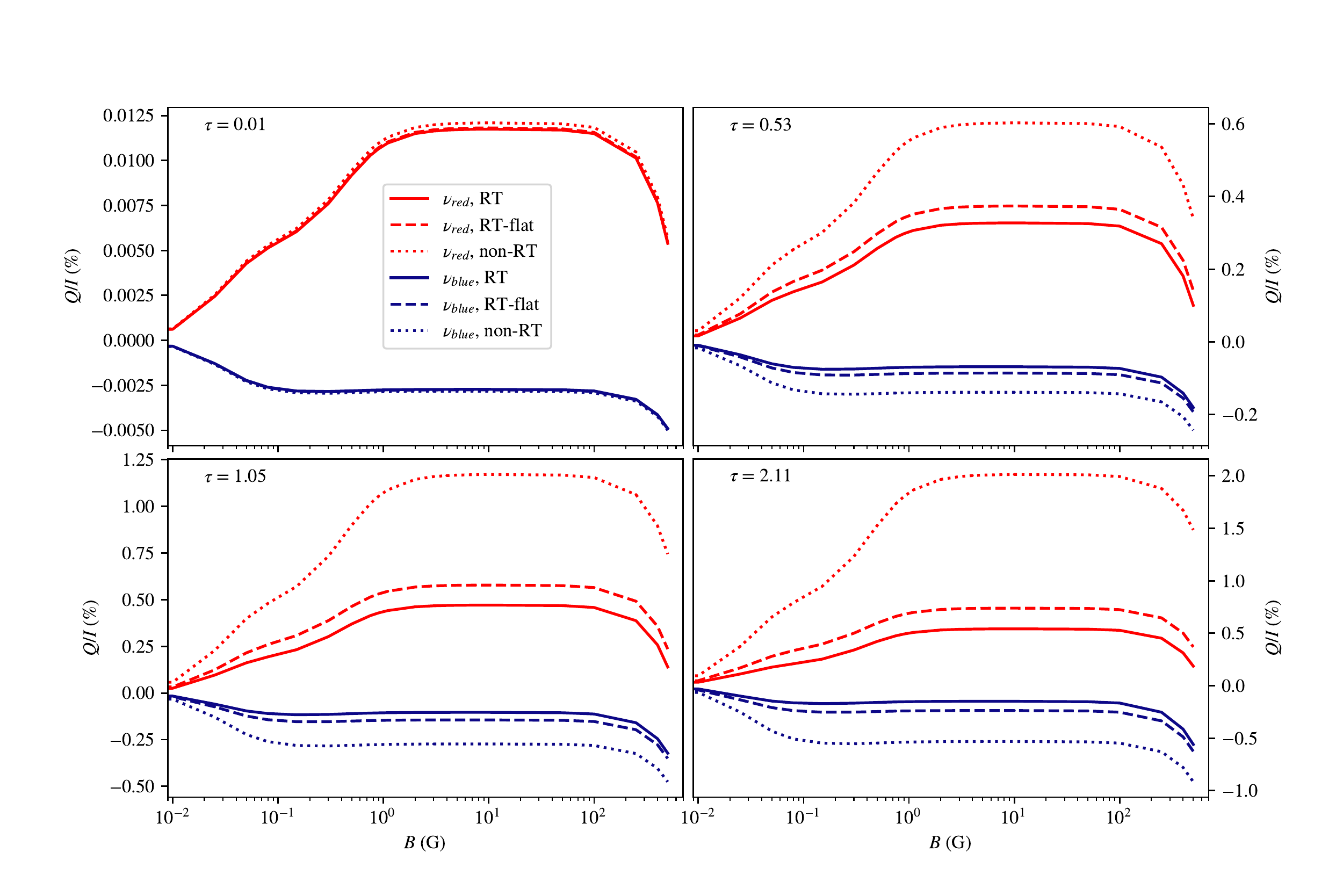}
\end{center}
\caption{Fractional linear polarization $Q/I$ as a function of magnetic field strength for different optical depths specified at the top-left of each panel. The blue (red) curves correspond to the blue (red) component of the 10830 multiplet. The dotted curve shows the result with the non-RT model, the dashed curve shows the result with the RT-flat model, and the solid curve shows the result with the RT model. Note that the vertical-axis scale is different in each panel.
}
\label{fig:B_tau_errors}
\end{figure*}

These findings are consistent with what is found in the observation and analysis of
the 10830 multiplet in prominences and filaments. In general terms, as explained above, it can
explain why prominences seem to show magnetic fields in the saturation regime, while quiet Sun
filaments seem to show magnetic fields in the fractions or units of gauss, i.e., in the Hanle
regime \cite[see][and references therein]{2022ARA&A..60..415T}. Regarding the
particular case of filaments, there are, as far as we know, only a few filament
observations with the dichroic blue component signal, and the inversion is usually
unable to achieve a completely satisfactory fit. \cite{TrujilloBuenoetal2002} show the fit to a
filament profile, demonstrating the dichroic origin of the blue component's linear
polarizaiton signal; however, their fit does not match the blue component's intensity
(see their Fig.\,4). \cite{Laggetal2004} show a fit to a profile in a flux emergence
region, neglecting lower level polarization, unable to fit the blue component's linear
polarization (see their Fig.\,4 and 5). Later, \cite{2008ApJ...683..542A} showed a fit
to the same profiles including lower level polarization and, while it is improved,
they are unable to simultaneously fit the linear polarization amplitude of both the blue
and red components (see their Fig.\,16, as well as Fig.\,14 for an example with other
filament observation). \cite{2019A&A...625A.128D,2019A&A...625A.129D} found and
studied this impossibility in the fitting, albeit their observations show peculiar
polarization signals with the same sign in both red and blue components, which require additional complexity
in the physical model beyond RT\footnote{We have not been able to find emergent Stokes profiles
with the same sign signals with our code in a constant property slab, except for
particular combinations of magnetic fields and velocities resulting in amplitudes much
smaller than those observed}. Other filament observations include those of
\cite{Kuckeinetal2012} and \cite{Xuetal2012}, who found mostly Zeeman linear
polarization profiles in filaments on top of active regions and thus
could be fit (see Figs.\,7-9 in \citealt{Kuckeinetal2012} and Fig.\,4 in
\citealt{Xuetal2012}). Curiously, \cite{Kuckeinetal2009} posit that a reduction
factor of $0.2$ was needed in the anisotropy in order to explain their filament
observations, a similar reduction to the one we find in our model for the red component's
anisotropy at $\tau=1$ ($\sim0.17$, see \fref{fig:radiation}).
All in all, observations of filaments in the 10830 multiplet
are relatively scarce, but they tend to show that the non-RT model may not be
enough to find a satisfactory fit to the observed linear polarization profiles.
Consequently, in order to correctly interpret observations in filaments with
relatively large optical depth, we think that RT must be taken into account,
the flat spectrum approximation must be relaxed, and that a model more complex than
a constant property slab is necessary to explain observations
such as those by \cite{2019A&A...625A.128D,2019A&A...625A.129D}.

\section{Conclusions\label{sec:concl}}

We have investigated the impact of RT effects on the polarization of the 10830 multiplet.
In particular, we have studied the effect that relaxing the flat spectrum approximation,
which requires to consider that both components are illuminated by identical radiation
fields, has in the anisotropy and in the emergent linear scattering polarization. To this
end we have modified the SEE for the multi-term atom, neglecting quantum interference terms
between the blue component's upper level and the two red component's upper levels.
Moreover, instead of calculating a singular average radiation field for the whole multiplet
(Eq.\,\ref{eq:jkq}) we compute the equivalent quantity for each of the components, where
the absorption profile is constructed by including only contributions to the absorptivity
from magnetic line components pertaining to each of the blue or red components
of the 10830 multiplet. We have implemented this modified set
of equations into a 1D RT code and calculated the 10830 multiplet emergent Stokes profiles
in a constant property slab model in order to compare our results with those obtained with
the usual modeling assumptions of these lines, namely, no RT and flat-spectrum across the
whole multiplet (non-RT model).

In the optical thin limit, our self-consistent calculations coincide with those of the
non-RT model, as expected. As we increase
the optical depth of the slab, the results start to diverge, and for not too large
optical depth we start observing significant differences in the
results. First, allowing for RT within the slab significantly affects the radiation
field. The mean intensity in the top (bottom) region of the slab shows a significant
defect (excess) in the lines due to the absorption (emission) by the rest of the slab
below (above). The radiation field anisotropy is instead diminished with respect to the
non-RT case due to the significant contribution of the radiation propagating along
the more inclined directions, which are optically thicker due to the geometry, as
anticipated by \cite{2007ApJ...655..642T}.

More important, especially for the diagnostic of the magnetic field, are the differences
in the emergent Stokes profiles. For a slab of optical depth $\tau=2$ we only find a
small difference in the red component's intensity, which could slightly impact the
determination of the optical depth and temperature (via the Doppler width).
However, the difference is remarkable for the linear scattering
polarization, of the order of a factor three in the signal of both components. Due to the larger
anisotropy in the non-RT model, diminished when horizontal RT within the slab
is accounted for, the linear polarization is significantly overestimated.
This can undoubtedly lead to an underestimation of the magnetic field or the filament height
for this particular disk-center filament configuration, a difference that can be
of orders of magnitude in the magnetic field strength in the worst-case scenarios.
In fact, filament observations showing scattering polarization signals
in the blue component, as far as we know, cannot be usually completely fit for both the blue
and red components. Moreover,
some observations show signals with the same sign in both the blue and red components,
something that we have not achieved to reproduce in a constant property slab even with RT,
what could mean that the constant property slab model (both RT and non-RT) is
too simplistic to model these observations.

The impact of the RT will be even more significant in full 3D geometry \citep[e.g.][]{2013PORTA}.
First, a non-homogeneous volume of plasma with enough optical depth will not only reduce the
anisotropy of the incoming radiation from the underlying disk, but also contribute to the
breaking of axial symmetry, an additional source of linear polarization which in the
commonly used slab model can only be accounted for by the magnetic field. Secondly, recent
3D NLTE RT calculations in an academic two-level atom model indicate that, already
for optical depths as small as $\tau=1$, RT plays an important role in spectral line formation
\citep[see Fig.\,9 of][]{2022A&A...659A.137S}.

Last but not least, we note that another important multiplet of \ion{He}{i}, namely the D$_3$
multiplet at 5876\,\AA, is most likely optically thin in all the structures of the outer solar
atmosphere and, therefore, it is less prone to be impacted by the effects discussed in this
paper. However, since the lower term of D$_3$ is the upper term of 10830, both multiplets are
coupled. Investigation on the impact of the 10830 transfer on the D$_3$ line remains one of
our research topics for the near future.

The results presented in this paper expose a potential problem of the simplified non-RT
modeling of such complex structures. While its simplifications allows for the implementation
of extremely fast inversion codes,
one needs to be careful when the plasma conditions are such that such simplifications are
not truly fulfilled. This emphasizes the relevance of developing novel diagnostic techniques
that can account for important physical ingredients such as the full three-dimensional
geometry and RT effects. For this reason, we are developing an inversion code based on the
ideas presented in \cite{2022A&A...659A.137S}, and will continue the investigation started
in this paper with significantly more generality (with 3D RT) in a forthcoming publication.


\begin{acknowledgements}
We are grateful to Luca Belluzzi, Javier Trujillo Bueno, and Andr\'es Asensio Ramos for number of suggestions that helped us to improve the paper.
We acknowledge the funding received from the European Research Council (ERC) under the European
Union's Horizon 2020 research and innovation program (ERC Advanced Grant agreement No\,742265).
J.\v{S}. acknowledges the financial support of the grant \mbox{19-20632S} of the Czech Grant
Foundation (GAČR) and the support from project \mbox{RVO:67985815} of the Astronomical Institute
of the Czech Academy of Sciences. T.P.A.'s participation in the publication is part of the Project
RYC2021-034006-I, funded by MICIN/AEI/10.13039/501100011033, and the European Union
“NextGenerationEU”/RTRP. M.J.M.G.'s participation in this research has been supported by the
project PGC-2018-102108-B-100 of the Spanish Ministry of Science and Innovation and by the
financial support through the Ramón y Cajal fellowship.
We also acknowledge the community effort devoted to the development of the following open-source packages that were used in this work: \texttt{numpy} \citep[\texttt{numpy.org},][]{numpy20}, \texttt{matplotlib} \citep[\texttt{matplotlib.org},][]{matplotlib}.
We made the code publicly available in a github repository \href{https://github.com/andreuva/He_1083_RT}{andreuva/He\_1083\_RT}.
\end{acknowledgements}

\bibliographystyle{aa}
\bibliography{ms.bib}

\end{document}